\documentclass[12pt,a4paper]{article}
\usepackage{graphicx,fancyhdr}
\usepackage{cite,citesort}
\begin{document}

\pagenumbering{arabic}

\vspace{-40mm}

\begin{center}
\Large\textbf{Criterion of multi-switching stability for magnetic
nanoparticles}\normalsize \vspace{8mm}

\large F. Porrati and M. Huth

\vspace{5mm}

\small\textit{Physikalisches Institut, J. W. Goethe-Universit\"at,
Max-von-Laue-Str. 1, D-60438 Frankfurt am Main, Germany}

\end{center}

\vspace{0cm}
\begin{center}
\large\textbf{Abstract}\normalsize
\end{center}

We present a procedure to study the switching and the stability of
an array of magnetic nanoparticles in the dynamical regime. The
procedure leads to the criterion of multi-switching stability to
be satisfied in order to have stable switching. The criterion is
used to compare various magnetic-field-induced switching schemes,
either present in the literature or suggested in the present work.
In particular, we perform micromagnetic simulations to study the
magnetization trajectories and the stability of the magnetization
after switching for nanoparticles of elliptical shape. We evaluate
the stability of the switching as a function of the thickness of
the particles and the rise and fall times of the magnetic pulses,
both at zero and room temperature. Furthermore, we investigate the
role of the dipolar interaction and its influence on the various
switching schemes. We find that the criterion of multi-switching
stability can be satisfied at room temperature and in the presence
of dipolar interactions for pulses shaped according to CMOS
specifications, for switching rates in the GHz regime.

\newpage

\begin{center}
\large\textbf{I. Introduction}\normalsize
\end{center}

The magnetization reversal and the state stability of magnetic
particles are key issues in magnetic technology applications. In
particles with uniaxial anisotropy the direction of the
magnetization can be changed by applying a field pulse either
parallel or perpendicular to the magnetic easy axis. The reversal
times, which depend on the speed and the direction of the applied
field, vary from many nanoseconds (damping switching) to hundreds
of picoseconds (ballistic precessional
switching)\cite{miltat,bertotti,bauer,kaka}. The stability of the
particle's magnetization configuration after switching depends on
several factors, such as the temperature of the system, the size
of the particle and the shape of the applied fields. Furthermore,
in arrays of magnetic particles the dipolar interaction and the
not perfectly equal shape of the particles shrinks the switching
margin used to switch a selected particle without affecting other
ones, the so-called half selected particles\cite{maffitt}.

In this work we present a procedure to study the stability of an
array of magnetic particles for magnetic-field-induced switching
schemes: First we analyze the stability of the half-selected
particles with respect to a chosen switching scheme; second, we
investigate the precessional switching of the particle of interest
(fully-selected particle)(section II-B); third, we check the
stability of the state obtained after switching as well as the
possibility of stable successive switchings. Such a procedure
leads us to define the \textit{criterion of multi-switching
stability} to be satisfied in order to have stable and repeatable
switching (sec. II-C). In the second part of the paper we extend
the investigation by considering the array as a function of
various parameters such as the thickness of the particles, the
duration of the pulse, the temperature (sec. III) and the strength
of the dipolar interaction (sec. IV).

\begin{center}
\large\textbf{II. Magnetization dynamics}\normalsize
\end{center}

\begin{center}
\large\textbf{A. Addressing schemes}\normalsize
\end{center}

\begin{figure}\center{\includegraphics[width=10cm]{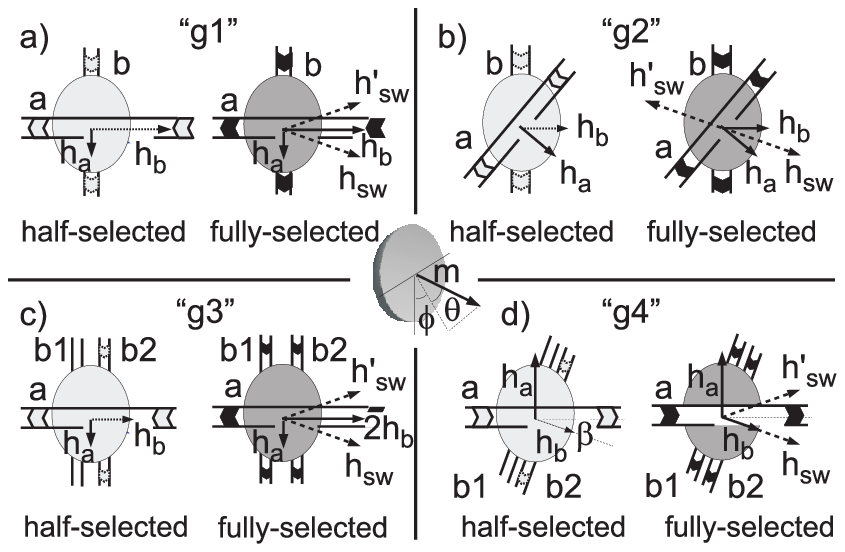}}
\caption{Addressing schemes. "g1": Classical cross-wire geometry.
"g2": Geometry after Ref~\cite{hillebrands}. "g3": Classical
geometry with improved stability. "g4": Tilted geometry with
improved stability. For each geometry we show the half-selected
particle and the fully-selected particle. To the half
(fully)-selected particle are applied either (both) the magnetic
field $h_a$ or (and) the magnetic field $h_b$.} \label{schemes}
\end{figure}

\begin{figure}\center{\includegraphics[width=14cm]{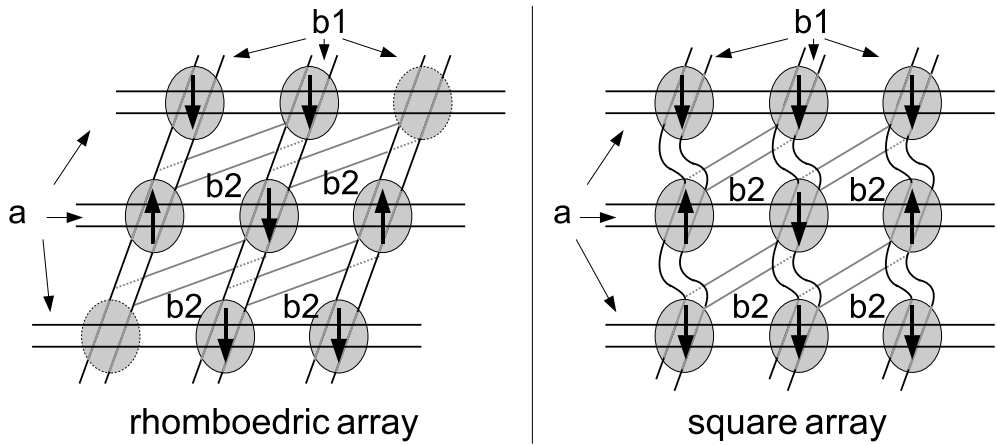}}
\caption{Array geometries. Left: rhomboedric array. Right: square
array. The wiring is similar in proximity of the magnetic
particles. Only the cylinders with arrows are used for the
calculation.} \label{wiring}
\end{figure}

In Fig.~\ref{schemes}a we draw the classical cross-wire geometry
used to write a magnetic cell by means of magnetic
fields\cite{maffitt}. The field
\textbf{\textit{h${_\textbf{sw}}$}}
(\textbf{\textit{h$^{\prime}{_\textbf{sw}}$}}) is used to switch
the magnetization from up to down (down to up) and it is given by
the sum of the fields \textbf{\textit{h${_\textbf{a}}$}} and
\textbf{\textit{h${_\textbf{b}}$}} produced by the current flow in
line \textit{a} (or "bit line") and line \textit{b} (or "word
line"), respectively. Such a scheme involves a bipolar current
flow in line \textit{a} necessary to obtain the two opposite
directions of \textbf{\textit{h${_\textbf{a}}$}}. In general, in
order to guarantee the stability of the half-selected particles,
the field components $h_{a}$ and $h_{b}$ must not exceed the
values specified by the border of the corresponding magnetic
astroid. The operation margin associated to the above mentioned
classical geometry (called "g1", from here on) does not guarantee
sufficient stability in the presence of dipolar interaction and
thermal fluctuations\cite{maffitt}. Because of that a much more
stable switching scheme has been developed\cite{engel}, which led
to the first commercial 4-Mb magnetic-RAM device. As an
alternative, to enlarge the operation margin one can reduce the
values of the fields $h_{a}$ and/or $h_{b}$ without changing the
field \textbf{\textit{h${_\textbf{sw}}$}}. Vectorially this is
obtained by varying the angle between the line-\textit{a} and
line-\textit{b}\cite{hillebrands}, see Fig.~\ref{schemes}b. In
such a geometry, (labelled "g2") the field
\textbf{\textit{h$^{\prime}{_\textbf{sw}}$}} is obtained by
inverting the current flow direction in both lines, \textit{a} and
\textit{b}. Thus, the addressing scheme is of "double-dipolar"
type. Another way to enlarge the operation margin is to add a
third set of lines to the wiring geometry, thus reducing the
magnetic field acting on the half-selected particles. For example,
one can divide line-\textit{b} in two sub-lines "$b_{1}$" and
"$b_{2}$", see Fig.~\ref{schemes}c-d, each one producing a
magnetic field \textbf{\textit{h${_\textbf{b}}$}}. Lines "$b_{1}$"
and "$b_{2}$" can be built in stack separated by a insolating
layer. These lines contribute to the switching process of a
fully-selected particle with the total magnetic field
2\textbf{\textit{h${_\textbf{b}}$}}. Due to the wiring geometry,
see Fig.~\ref{wiring}, a half-selected particle is subjected only
to the field \textbf{\textit{h${_\textbf{b}}$}}, which enhances
its stability. In the geometry of Fig.~\ref{schemes}c ("g3") lines
"$b_{1}$" and "$b_{2}$" are parallel to the magnetic easy axis of
the particle, as for the geometry "g1". The switching scheme in
the two cases is similar. In the geometry of Fig.~\ref{schemes}d
("g4") lines "$b_{1}$" and "$b_{2}$" are tilted by the  angle
$\beta$ with respect to the magnetic easy axis. The magnetic
fields \textbf{\textit{h${_\textbf{a}}$}} and
\textbf{\textit{h${_\textbf{b}}$}} are applied to the
half-selected particles. In the fully-selected particle the
magnetization \textbf{\textit{m}} switches by application of the
magnetic field \textbf{\textit{h${_\textbf{sw}}$}} or
\textbf{\textit{h$^{\prime}{_\textbf{sw}}$}}, with
\textbf{\textit{h${_\textbf{sw}}$}}=2~\textbf{\textit{h${_\textbf{b}}$}}.
The fields ${h_a}$ and ${h_{sw}}$ are related by the expression
${h_a}$=2~$h_{sw}$~sin~$\beta$ so that
\textbf{\textit{h${_\textbf{sw}}$}} and
\textbf{\textit{h$^{\prime}{_\textbf{sw}}$}} are equal in module
but symmetrically oriented with respect to the minor axis of the
ellipse. In contradistinction to the examples in
Figs.~\ref{schemes}a-c, this addressing scheme allows easy
switching by the use of unipolar magnetic field pulses, as will be
detailed later.

\begin{center}
\large\textbf{B. Stability regions, switching trajectories and
pulse synchronism}\normalsize
\end{center}

In the following section we study the switching properties and the
stability of an isolated elliptical cylinder. To be explicit, the
major and the minor semi-axis of the cylinder are 50~nm and 40~nm
long, respectively. The hight is 1~nm. The direction of the
average magnetization is given by the latitude $\theta$, measured
from the plane of the particle, and the azimuthal angle $\phi$,
measured from the easy axis of the particle (Fig.~\ref{schemes}).
The material is iron with cubic anisotropy constant
k=1.5$\times10^{4}$~J/m$^{3}$\cite{brockmann}, exchange stiffness
constant A=2.1$\times10^{-11}$~J/m and saturation magnetization
M$_{s}$=1.714$\times10^{6}$~A/m$^{3}$\cite{scheinfein}. For this
choice of the geometry and the magnetic constants the total energy
is dominated by the stray field energy, being more than one order
of magnitude larger than the exchange energy and the anisotropy
energy. Thus, the system has two minima: one located at
($\theta=0^{\circ}$, $\phi=0^{\circ}$), the "down" state or bit
"0", the other at ($\theta=0^{\circ}$, $\phi=180^{\circ}$), the
"up" state or bit "1". Such an example is representative of any
bi-stable system for which the energy of the particle is dominated
by its shape.

\begin{figure}\center{\includegraphics[width=10cm]{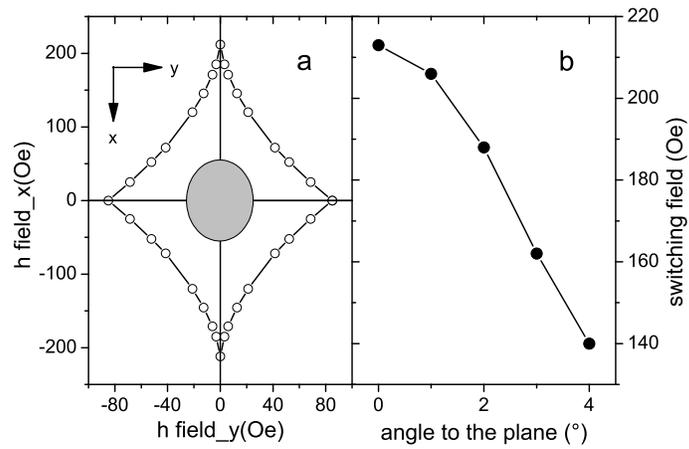}}
\caption{a: Dynamical astroid. The diagram is obtained for an
elliptical cylinder with semi-axis of 50 nm and 40 nm, hight 1 nm.
The field pulse is a step function with 200 ps rise-time. The
damping factor is $\alpha$=0.02. b: Switching field vs. angle to
the plane $\theta$.} \label{astroid+z_switching}
\end{figure}

We perform micromagnetic simulations\cite{scheinfein} in order to
study the magnetization trajectories and the stability of the
magnetic states obtained after application of an external magnetic
field pulse. The Landau-Lifshitz-Gilbert (LLG) equation of motion
is solved in the dynamic regime by using a value of the damping
factor $\alpha$=~0.02. The simulation volume is divided in unit
cells of parallelepiped shape with an edge of 2~nm in the plane
and 1~nm in thickness. In Fig.~\ref{astroid+z_switching}a we show
the dynamical astroid\cite{miltat} obtained for a pulse with a
rise time of 200~ps. By applying a magnetic field antiparallel to
the magnetization direction we find that the field necessary to
overcome the barrier between the two minima is 213~Oe. It is worth
to note that often in the literature the analysis of the switching
behavior of a magnetic particle is 2-dimensional, while the
complete scenario is given by means of a 3D
analysis\cite{thiaville}. To understand this fact, in
Fig.~\ref{astroid+z_switching}b we plot the magnetic field
necessary to switch the magnetization as a function of the
latitude $\theta$. For the selected $\alpha$=~0.02 the switching
field is reduced dramatically by 73~Oe when increasing the
latitude to only 4${^\circ}$. The corresponding reduction in the
damping regime (i.e. $\alpha$=~1) is 8~Oe. This difference is due
to the differing amplitudes of the damping oscillations, viz. to
the torque produced by the applied field on the average
magnetization. This result shows that a 3D analysis is fundamental
to study the switching and the stability of magnetic
nanoparticles, in particular in the dynamical regime.

\begin{figure}\center{\includegraphics[width=10cm]{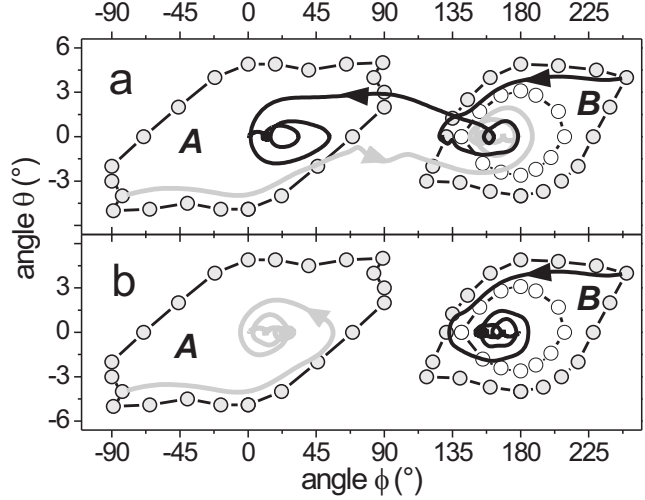}}
\caption{Criterion of multi-switching stability. The regions
\textit{A} and \textit{B} contain the coordinates for which the
magnetization is stable under the action of one single field,
namely \textbf{\textit{h${_\textbf{a}}$}} or
\textbf{\textit{h${_\textbf{b}}$}}, respectively. The regions
enclosed by the grey (open) dots are obtained for the
[90$^\circ$-$\beta$]-cross-wire (90$^\circ$-cross-wire) geometry
for \textit{h${_{sw}}$}=100~Oe (\textit{h${_{sw}}$}=80~Oe). Panel
a: examples of switching (non perfectly synchronized pulses). The
grey (black) line represents the trajectory of the average
magnetization from the boundary of region \textit{A} (\textit{B})
into region \textit{B} (\textit{A}) by means of the field
 \textbf{\textit{h$^{\prime}{_\textbf{sw}}$}}
(\textbf{\textit{h${_\textbf{sw}}$}}).  Panel b: examples of no
switching (non-perfectly synchronized pulses). The grey (black)
line represents the trajectory from the boundary of region
\textit{A} (\textit{B}) into the same region by means of the field
\textbf{\textit{h${_\textbf{sw}}$}}
(\textbf{\textit{h$^{\prime}{_\textbf{sw}}$}}).} \label{stability}
\end{figure}

In a 3D picture, in order to prove the stability of half selected
particles we define a region in the plane ($\phi$~,~$\theta$)
where the magnetization is stable under the action of one single
pulse. Let us consider a bi-stable system with energy minima in
the "down" state and in the "up" state. If the magnetization is
"down" ("up") it may switch because of field
\textbf{\textit{h${_\textbf{a}}$}}
(\textbf{\textit{h${_\textbf{b}}$}}), according to the addressing
scheme geometry "g4". Therefore we define two regions of stability
in the plane ($\phi$~,~$\theta$): region \textit{A}, where the
magnetization is stable under the application of the field
\textbf{\textit{h${_\textbf{a}}$}} and region \textit{B}, where
the magnetization is stable under the application of the field
\textbf{\textit{h${_\textbf{b}}$}} (see Fig.~\ref{stability}). The
stability region \textit{A} is obtained by applying a step
function of module \textit{h$_a$}= 68.4~Oe with 200~ps rise time
(the value of \textit{h$_a$} is related to \textit{h$_{sw}$}=
100~Oe, using $\beta=20^{\circ}$). At $\theta=0^{\circ}$ the
magnetic state is stable for $-69^{\circ}\leq\phi\leq69^{\circ}$.
At $\phi=0^{\circ}$ the magnetic state is stable for
$-4.9^{\circ}\leq\theta\leq4.9^{\circ}$. A similar analysis is
done for the stability region \textit{B}. This region is obtained
by applying a step function of module \textit{h$_b$}= 50~Oe with
200~ps rise time. At $\theta=0^{\circ}$ the magnetic state is
stable for $131^{\circ}\leq\phi\leq228^{\circ}$. At
$\phi=180^{\circ}$ the magnetic state is stable for
$-4^{\circ}\leq\theta\leq4.9^{\circ}$. By analyzing the two
stability regions we notice that, on the one hand, the stability
interval is quite large if the average magnetization lies in- or
almost in-plane; on the other hand, the stability interval is
strongly reduced for $\theta\neq0$. As a consequence only the
particles with the magnetization close to the plane are stable.
Note that in order to have stability in one region the
magnetization has to relax from each border-point of the stability
region into the minimum with a damping trajectory belonging to
said region. In that case, a pulse applied at any time cannot lead
to the instability of a half-selected particle. In
Fig.~\ref{stability} and in the following we plot stability
regions obtained after a refinement procedure, which consists in
reducing the stability regions until the relaxing trajectories
fully belong to the respective regions.

\begin{figure}\center{\includegraphics[width=10cm]{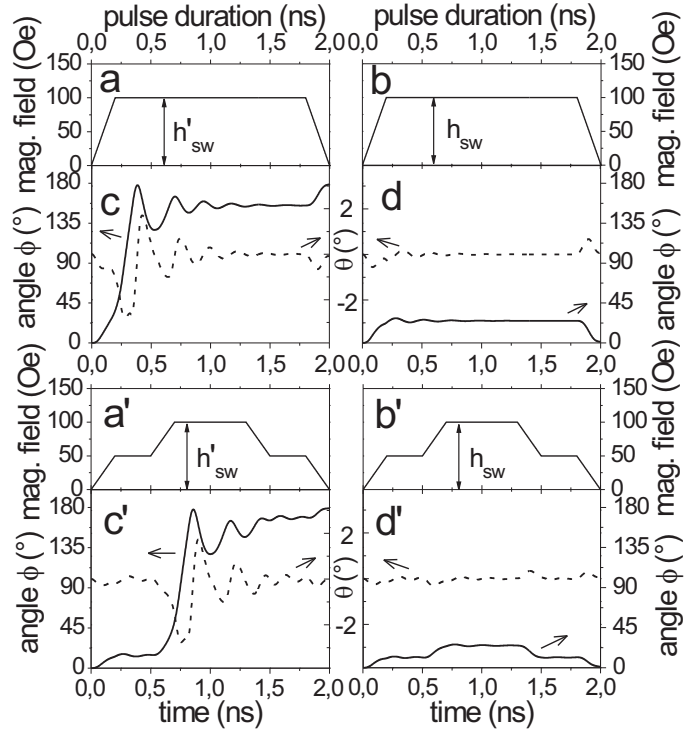}}
\caption{Switching of the magnetization. Panels a-b (a'-b'): shape
of the magnetic field with synchronized (non-perfectly
synchronized) pulses. Panels c-d/c'-d': magnetization
trajectories. Panel c-c': The magnetization switches from "down"
to "up" by application of the field
\textbf{\textit{h$^{\prime}{_\textbf{sw}}$}}. Panel d-d': The
magnetization does not switch by application of the field
\textbf{\textit{h${_\textbf{sw}}$}}.} \label{switch}
\end{figure}

The study of the magnetization switching is done by using the
pulses shown in Fig.~\ref{switch}. The shape of the pulse in
Fig.~\ref{switch}a-b is "trapezoidal". This pulse is characterized
by the four segments ($t_p, t_r, t_{pl}, t_f$), $t_p$ being the
duration of the pulse, $t_r$ the rise time, $t_{pl}$ the central
plateau and $t_f$ the fall time. Each time is given in
nanoseconds. For example, the pulse in Fig.~\ref{switch}a-b is
characterized by (2, 0.2, 1.6, 0.2). The direction of the magnetic
field is \textbf{\textit{h$^{\prime}{_\textbf{sw}}$}} and
\textbf{\textit{h${_\textbf{sw}}$}}, respectively, with maximum
module $h_m$=100~Oe. In Fig.~\ref{switch}~c-d we plot the response
of the averaged magnetization to the field pulses. In both cases
the initial state of the system is "down", while the final state
is either "up" or "down". By means of the field
\textbf{\textit{h$^{\prime}{_\textbf{sw}}$}} the magnetization
rotates towards the "up" state. The switching takes place in the
first 380~ps by means of a half precession. After that, the
magnetization shows damped oscillations in a transient energy
minimum at ($\theta\approx0^{\circ}$, $\phi\approx154^{\circ}$).
In the last 200~ps, the magnetization reaches the new minimum in
absence of the applied field. By means of the field
\textbf{\textit{h${_\textbf{sw}}$}} the averaged magnetization
does not switch. In the first 200~ps the magnetization reaches a
transient temporary minimum at ($\theta\approx0^{\circ}$,
$\phi\approx24^{\circ}$). The initial minimum is established after
pulse termination. For both, switching and no switching, the final
states are stable to the application of successive single magnetic
field pulses since the magnetization trajectories terminate inside
the regions of stability. The same conclusion is valid for the
transient energy minima. As a consequence, stable switching is
possible also for shorter pulses. Note that since the shape of the
particle is not an ellipsoid the magnetization does not rotate
coherently. Rather, due to the small size of the particle the
magnetization rotates quasi-uniformly.

The pulses used above to switch the magnetization are given by the
vectorial sum of two sub-pulses
\textbf{\textit{h${_\textbf{sw}}$}}=
2\textbf{\textit{h${_\textbf{b}}$}} and
\textbf{\textit{h$^{\prime}{_\textbf{sw}}$}}=\textbf{\textit{h${_\textbf{sw}}$}}
+\textbf{\textit{h${_\textbf{a}}$}}. Up to here we have considered
the sub-pulses as synchronized. Experimentally such a condition is
difficult to realize in a large array of magnetic
particles\cite{maunoury}. In order to study the effect of the
non-perfect synchronism of the sub-pulses we consider the "double
trapezoidal" pulse shown in Fig.~\ref{switch}~a'-b'. This pulse is
characterized by five segments ($t_p, t_r, t_{pl1}, t_{pl2},
t_f$), $t_p$ being the duration of the pulse, $t_r$ the rise time
from zero to the first plateau and from the first plateau to the
central one, $t_{pl1}$ the first and the last plateaus, $t_{pl2}$
the central plateau, $t_f$ the fall time from the central to the
last plateau and from the last plateau to zero. Therefore, the
pulse in Fig.~\ref{switch}a'-b' is characterized by (2, 0.2, 0,3,
0.6, 0.2). In Fig.~\ref{switch}~c'-d' we plot the response of the
average magnetization to such a field pulse. As above, in both
cases the initial state of the system is "down". The effect of the
non-perfect synchronism is evident by comparing
Fig.~\ref{switch}~c'-d' with Fig.~\ref{switch}~c-d. The main
effect on the switching is the time shift from 380~ps to 850~ps in
which the magnetization rotates towards the new energy minimum
orientation (compare Figs.~\ref{switch}~c/c'). The non-perfect
synchronism affects also the no-switching process. As shown in
Fig.~\ref{switch}~d', the direction of the magnetization follows
the shape of the pulse. The maximum value of the angle $\phi$ is
reached at $t\approx0.7$~ns, with a shift of $\approx0.5$~ns with
respect to Fig.~\ref{switch}~d. For both, switching and
no-switching, the final states are stable, as in the case of
synchronized pulses. Since the effect of the non-perfect
synchronism is to shift the switching time without affecting the
stability of the final state, the time scale of the switching will
depend directly on the experimentally achievable timing. The set
of switching processes analyzed above is completed by the two
symmetric switchings for which the magnetization is initially
oriented in the "up" direction. If the four switchings are
successful, the following proposition is valid: whatever is the
initial state, the field
\textbf{\textit{h$^{\prime}{_\textbf{sw}}$}}
(\textbf{\textit{h${_\textbf{sw}}$}}) causes the magnetic particle
to end up in the "up" ("down") state. Such a property allows
successive magnetization reversals without time delay since no
knowledge of the magnetization direction is required, in
contradistinction to toggle switching schemes \cite{schumacher2}.

\begin{center}
\large\textbf{C. Criterion of multi-switching
stability}\normalsize
\end{center}

Until this point we have studied the response of a magnetic
particle to a field pulse by considering the initial magnetization
as being located in one of the energy minima of the system. In the
following we extend the analysis to a magnetic particle with
initial state represented by any of the points inside the
stability regions \textit{A} and \textit{B}. As above we consider
the switching and the no-switching processes. Each point in
regions \textit{A} and \textit{B} is tested with regard to the
field pulses \textbf{\textit{h$^{\prime}{_\textbf{sw}}$}} and
\textbf{\textit{h${_\textbf{sw}}$}}. If all the magnetization
trajectories converge inside the corresponding stability regions
the system is considered stable to successive switchings. If these
convergence conditions are satisfied, we say that the system
satisfies the \textit{criterion of multi-switching stability}. In
Fig.~\ref{stability}a-b we plot some examples of switching and
no-switching trajectories obtained with the magnetic pulse shown
in Fig.~\ref{switch}a'-b'. The initial states of the magnetization
are located at the border of the stability regions. After pulse
application the magnetization settles inside regions \textit{A} or
\textit{B}. Such a behavior is found for all the points located at
the border of the stability regions, which is sufficient to prove
the convergence of any initial state belonging to the stability
regions. By means of such a procedure we find that the criterion
of multi-switching stability is satisfied by using both,
synchronized and non-perfectly synchronized pulses.

The analysis presented in this section, valid for the switching
scheme "g4", can be extended to any other switching geometry. In
sec. IV-B we present a comparison between the addressing schemes
introduced in sec. II-A.

\begin{center}
\large\textbf{III. Temperature, thickness and pulse rise-time
dependencies}\normalsize
\end{center}

In the former sections we have used the LLG equation of motion,
which provides a zero-temperature description of the magnetization
processes. A non-zero-temperature description is obtained by means
of the stochastic LLG equation, based on
Langevin-dynamics\cite{garcia}. In this section we extends the
former analysis to the room-temperature regime. Furthermore, we
study the switching and the stability as a function of the
thickness of the elliptical cylinder and the rise time of the
magnetic field pulse.

\begin{figure}\center{\includegraphics[width=10cm]{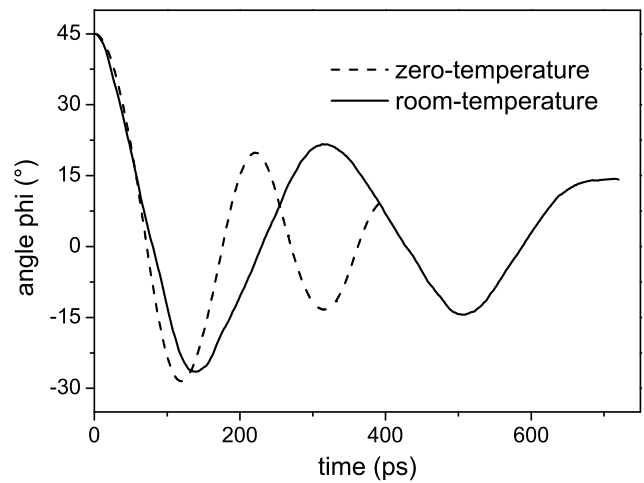}}
\caption{Damped magnetization oscillations vs temperature. The
average magnetization relaxes more rapidly at zero temperature
than at room temperature. This affects the amplitude of the
regions of stability, see Fig.\ref{comparison_stability}.}
\label{oscillations}
\end{figure}

\begin{figure}\center{\includegraphics[width=10cm]{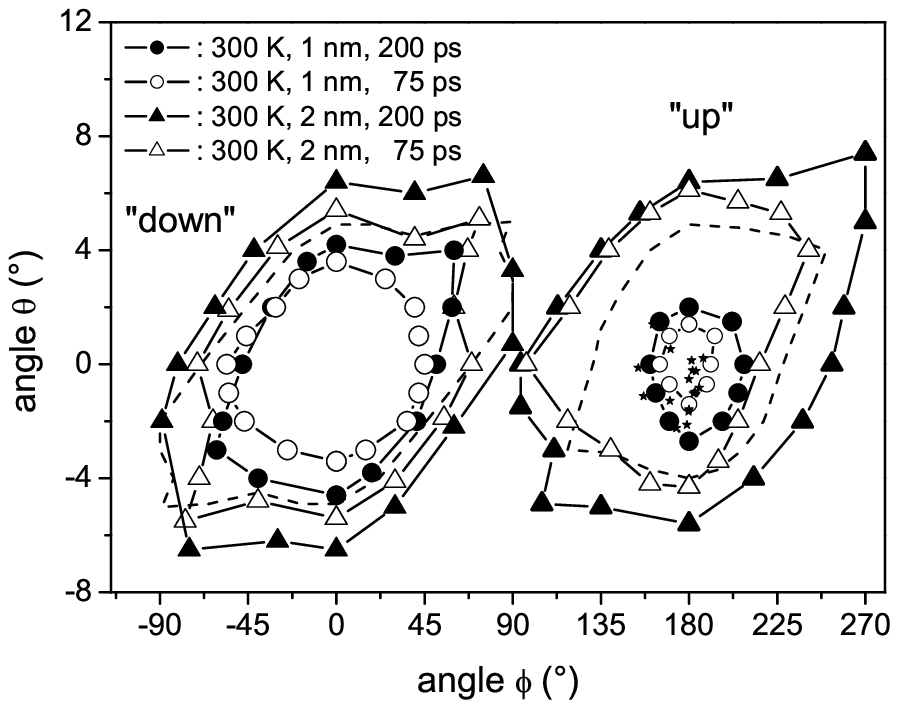}}
\caption{Temperature and rise time dependencies of the stability
regions.\newline Temperature dependence: compare dashed line (zero
temperature) with full-dot line (room temperature).\newline Rise
time dependence: compare full-dot/triangle lines ($t_r$=200~ps)
with open-dot/triangle lines ($t_r$=75~ps).\newline Particle
thickness dependence: compare open/close-dot lines (1~nm
thickness) with open/close-triangle lines (2~nm thick).\newline
Star-points: final states obtained after switching from the "down"
stability region (open-triangles) for a pulse with duration
t=700~ps.} \label{comparison_stability}\end{figure}

In section II-B we have shown that for a cylinder with hight
$t=1$~nm at zero temperature for a pulse with rise time
$t_r$=200~ps the switching field is $h_{sw}=213$~Oe. In
correspondence, at room-temperature ($300$~K) we find
$h_{sw}=162$~Oe. This result is in accordance with the reduction
of the coercivity with the temperature found in arrays of
submicron elliptically patterned permalloy thin films\cite{deak}.
In order to understand the reduction, in Fig.~\ref{oscillations}
we plot the time dependence of the azimuthal angle $\phi$ at zero
and room temperature, respectively. The plot is obtained with the
magnetization starting conditions $\theta=0^\circ$,
$\phi=45^\circ$ in zero field. One observes that at zero
temperature the first two oscillations take place within 400~ps,
while at room temperature this happens within 710~ps. Therefore,
the duration of the oscillations increases with temperature.
Because long oscillations are associated to a long action of the
torque between the magnetization and the applied field, in order
to switch the magnetization at room temperature it is necessary to
apply a smaller field than at zero temperature. For this same
reason in Fig.~\ref{comparison_stability} the regions of stability
at zero temperature (dashed line)  are larger than those at room
temperature (full-dot line). The regions are calculated employing
the switching scheme "g4", following the procedure of sec.~II-B.
Furthermore, since at room temperature the system is described in
the framework of the Langevin dynamics, the stability of the
regions against stochastic fluctuations is checked by considering
each border point as stable only after three "destabilization"
attempts have led to the same convergence point. The analysis of
the damped oscillations explains also the observed growth of the
stability regions with increasing the rise time from $t_r$=75~ps
(open-dot~line) to $t_r$=200~ps (full-dot~line). In fact, a pulse
with short rise time exerts its influence already on the first
magnetization oscillation, which is the most unstable because it
has the largest amplitude.

After the calculation of the stability regions we now study the
switching of the magnetization. We consider pulses with durations
of 1~ns, 1.5~ns and 2~ns, with rise times of 75~ps and 200~ps at
room temperature. The amplitude of the pulse is $h_{sw}=100$~Oe.
Comparing the results for all the parameters, when switching from
the "down" stability region some of the trajectories do not
terminate inside the "up" stability region. Therefore the
criterion of multi-switching stability is not satisfied. One way
to satisfy the criterion is to choose a long enough pulse and fall
time, so to allow the magnetization to relax inside the stability
region. However, this method does not improve the intrinsic
stability of the system, represented by the area of the stability
region. In the presence of the dipolar interaction this area is
even further reduced (see sec. IV) and thus the
switching-stability criterion will not be satisfied, even for
longer pulses. Another way to enlarge the stability region is to
increase the thickness of the particle, which also causes an
increase of the coercivity\cite{cowburn}. For an elliptical
cylinder with the same size as used above and height t=2~nm, we
obtain at zero temperature a switching field $h_{sw}=230$~Oe by
using a pulse with rise time $t_r$=200~ps directed antiparallel to
the magnetization direction and a damping factor $\alpha=0.02$. At
room temperature we obtain $h_{sw}=207$~Oe. In
Fig.~\ref{comparison_stability} we plot the stability regions
obtained at room temperature for $h_{sw}=130$~Oe. This amplitude
is higher than the one used for $t=1$~nm ($h_{sw}=100$~Oe), which
does not lead to stable-switching. The regions obtained for
thickness $t=2$~nm are larger than those obtained for $t=1$~nm,
especially much so for the regions around the "up" state
($\phi=180^\circ$). As for the cylinder with $t=1$~nm, also for
$t=2$~nm the regions obtained for $t_r=75$~ps (open-triangle~line)
are smaller than those obtained for $t_r=200$~ps
(full-triangle~line).

\begin{figure}\center{\includegraphics[width=10cm]{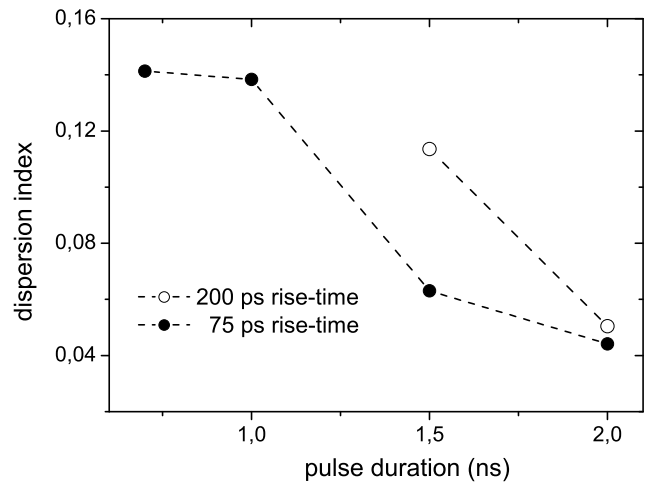}}
\caption{Dispersion index vs. pulse duration. The dispersion of
the final states decreases with decreasing the rise time and with
increasing pulse duration. See text for details.}
\label{dispersion_index}
\end{figure}

In the following we study the switching of the magnetization
employing scheme "g4" by applying the field $h_{sw}=130$~Oe to a
cylinder with $t=2$~nm at room temperature. We consider two pulses
characterized by (1.5, 0.2, 0.2, 0.3, 0.2) and (2, 0.2, 0.2, 0.8,
0.2). As explained in sec. II-B, this "double trapezoidal" pulse
shape takes into account the non-perfect synchronism between
sub-pulses. We observe that all the trajectories starting from the
border of the "down" stability region terminate inside the "up"
stability region. It is important to evaluate under which
switching conditions the final states lie in close proximity to
each other in the plane ($\theta-\phi$). This evaluation is then
an implicit measure of the stability of the system in the presence
of the dipolar interaction since in that case the stability
regions tend to shrink. In that sense we define the dispersion
index $\overline{d}=\sum_{i=1}^{N}d_{i}/N$ as the average distance
between the final magnetization states and their midpoint
($\overline{x}, \overline{y}, \overline{z}$), with
$d_i=((\overline{x}-x_{i})^{2}+(\overline{y}-y_{i})^{2}+(\overline{z}-z_{i})^{2})^{1/2}$,
$\overline{x}=\sum_{i=1}^{N}x_{i}/N$,
$\overline{y}=\sum_{i=1}^{N}y_{i}/N$,
$\overline{z}=\sum_{i=1}^{N}z_{i}/N$ and ($x_{i}$, $y_{i}$,
$z_{i}$) the cartesian coordinate of the final state \textit{i}.
The dispersion index related to the two pulses with duration
$t_p=1.5$~ns and $t_p=2$~ns are $\overline{d}=0.114$ and
$\overline{d}=0.05$, respectively. The dispersion is smaller for
the pulse with longer duration, see Fig.~\ref{dispersion_index}.
Next we consider two pulses characterized by (1.5, 0.075, 0.2,
0.8, 0.075) and (2, 0.075, 0.2, 1.3, 0.075). This is done in order
to study the influence of the rise time on the switching behavior
of the system. As in the former case, by switching from the "down"
stability region, all the trajectories terminate inside the "up"
stability region. The dispersion indices are $\overline{d}=0.063$
and $\overline{d}=0.044$, respectively. By comparing these values
with those obtained for $t_r=t_f=0.2$~ns, we find that a shorter
rise and fall time result in a smaller dispersion. Finally, we
consider two pulses characterized by (1, 0.075, 0.1, 0.5, 0.075)
and (0.7, 0.075, 0.1, 0.2, 0.075). In that case we want to study
the switching in the sub-nanosecond or nanosecond regime. To do
that we have to shorten the duration of the plateau between
sub-pulses. Again we have stable switching from the "down" to the
"up" stability regions. The dispersion indices are
$\overline{d}=0.141$ and $\overline{d}=0.138$, respectively.
Concluding, the largest dispersion is obtained for the shortest
pulse. For this case the final states are shown by the stars in
Fig.~\ref{comparison_stability}. It is important to note that for
all the pulses with different duration considered we have proved
that the criterion of multi-switching stability is satisfied.

\begin{center}
\large\textbf{IV. Dipolar interaction}\normalsize
\end{center}

\begin{figure}\center{\includegraphics[width=10cm]{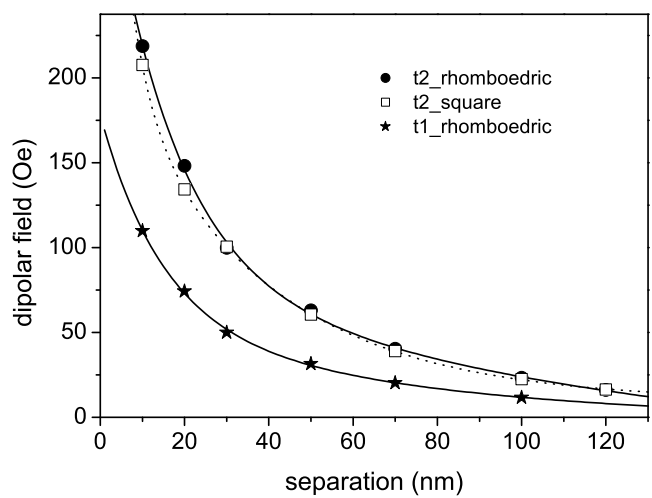}}
\caption{Average dipolar field vs. separation of the particles.
The value of the field decreases exponentially by increasing the
separation. The dipolar coupling is weaker for arrays of cylinder
with hight 1 nm (stars) than for arrays of cylinder with hight 2
nm (full dots). The array geometry, either rhomboedric or squared,
influence the strength of the dipolar coupling only for small
separation of the cylinders (compare full dots with open square).}
\label{dipolar_decay}
\end{figure}

Until this point we have studied the stability and the switching
behavior of a single particle. The results obtained changes for an
array of magnets, i.e. in the presence of the dipolar interaction.
It is known that for highly packed arrays the dipolar coupling can
modify the micromagnetic structure \cite{bromwich,fab1} and induce
transitions between the magnetic states of a
particle\cite{guslienko,fab2}. Furthermore, it influences the
reversal of the magnetization in the quasi-static
regime\cite{dunin,abraham,fab1}. In fast dynamics, the guideline
to study the precessional switching in the presence of the dipolar
interaction was given in Ref.\cite{devolder}. In particular, the
authors study the effect of the inter-particle interaction on the
operation margin in an array of cells like for a magnetic-ram
device. Recently, the effect of the dipolar interaction on the
dynamics of the precessional switching for a system of two coupled
ellipsoids has been investigated\cite{pham}. In both
cases\cite{devolder,pham} the analysis is performed at
zero-temperature in the framework of the single-spin
approximation. In the following we study the influence of the
dipolar interaction in an array of magnetic particles by means of
micromagnetic simulations. We consider two possible arrangements
of the particles: (see Fig.~\ref{wiring}) a rhomboedric array and
a square array. The wiring of the two geometries is slightly
different, although at the magnetic particle site the direction of
the lines "\textit{a}" and "\textit{b}" is identical. Each
magnetic particle is subject to a dipolar field which depends on
the magnetic configuration of the surrounding particles. After
symmetry considerations we distinguish among 32 and 30 different
magnetization arrangements for the rhomboedric and the square
geometries, respectively. In order to study the effect of the
dipolar interaction we consider the magnetic field produced on the
particle of interest by its first and second neighbors. The
contribution of particles of the third or higher coordination
shell is neglected. In Fig.~\ref{wiring} we show the configuration
that produces the largest value of the average dipolar field among
all the possible magnetic arrangements. Therefore this
configuration is the "key"-configuration to study the stability of
the system in the presence of the dipolar interaction. In
Fig.~\ref{dipolar_decay} we plot the average dipolar field as a
function of the distance "\hspace{-1 mm}\textit{s}" between the
particles. As can be expected, the dipolar coupling decreases with
increasing "\hspace{-1 mm}\textit{s}" and with decreasing
thickness of the magnetic particle. In particular we notice that
the geometry of the array does not play a crucial role, unless
$s\leq30$~nm.

\begin{center}
\large\textbf{A. Spatially- and time-dependent dipolar
coupling}\normalsize
\end{center}

\begin{figure}\center{\includegraphics[width=16cm]{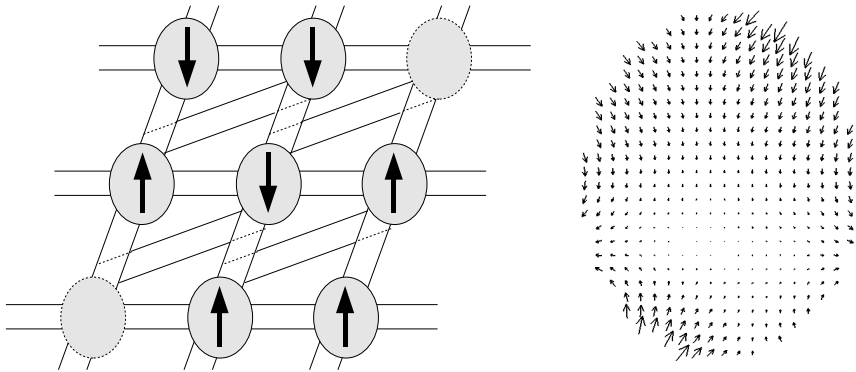}}
\caption{Dipolar configuration. Left: one of the possible magnetic
arrangement for the rhomboedric geometry. The dipolar field acting
on the central particle is calculated by means of its first and
second neighbors. Right: Resulting spatially dependent dipolar
field calculated on the central particle.}
\label{dipolar_configuration}\end{figure}

In general the dipolar field varies in space and is time
dependent, which was not considered in Ref~\cite{devolder}. In
Fig.~\ref{dipolar_configuration} we show one possible magnetic
configuration for the rhomboedric geometry and the corresponding
spatially dependent dipolar field of the central particle. The
dipolar field is spatially not uniform, both in strength and
direction. The simplest way to study the magneto-dynamics in an
array of particles is to consider the spatial average of the
dipolar field, without any time dependence. Note that such an
approximation is different from the single-spin
approximation\cite{devolder,pham} because the average field is
applied to a spatially extended micromagnetic entity. The most
complete way to study the magneto-dynamics in an array of
particles is to consider the effect of the magnetization rotation
of the particle of interest and the movement of the half-selected
particles on the dipolar field strength and direction. These
motions give rise to the time dependence of the dipolar field.

\begin{figure}\center{\includegraphics[width=10cm]{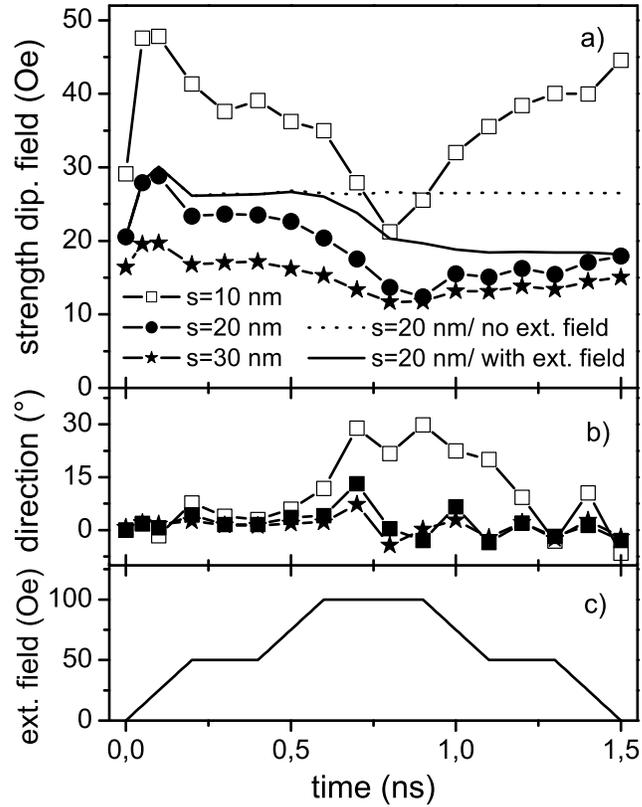}}
\caption{Time-dependent dipolar for various distances between the
particles. The field is calculated on the central particle for the
configuration of Fig.~\ref{dipolar_configuration} after
application of a pulse with duration $t_p$=1.5~ns. a): field
strength. b): field direction. c): pulse shape. Legend:
Open-square line for s=10~nm. Full-dot line for s=20~nm. Star line
for s=30~nm. Dot line (full line): for s=20~nm, dipolar field
calculated without (with) application of the external field
pulse.} \label{time-dependent_dipolar_field}
\end{figure}

In the following we study the magnetization process of a particle
according to the scheme "g4" and apply the field $H_{sw}=100$~Oe
to an elliptical cylinder with $t=1$~nm at zero temperature. The
pulse is characterized by (1.5, 0.2, 0.2, 0.8, 0.2). In
Fig~\ref{time-dependent_dipolar_field}a we plot the time
dependence of the dipolar field for the configuration in
Fig.~\ref{dipolar_configuration}. The open-square line depicts the
time-dependent dipolar field calculated for $s$=10~nm. The value
of the dipolar field at $t$=0 corresponds to the one used in the
average dipolar field approximation: $\overline{h_d}$=29~Oe, where
the direction of the sum vector is along the major axis of the
ellipse. The increase of the dipolar-field in the first 50~ps is
associated to the new minimum reached after relaxation from the
starting condition, where the magnetization is imposed to be
parallel to the major axis of the ellipse. After that, the
magnetization of the fully-selected particle starts to rotate.
Such a rotation modifies the orientation of the neighboring
particles leading to a decrease of the dipolar field. The minimum
is reached after 800~ps. At this moment the average magnetization
direction of the fully-selected particle is parallel to the minor
axis of the ellipse. Successively, the magnetization turns back
into the initial minimum without switching and the dipolar field
increases to the value taken before starting the rotation. The
full-dot line shows the time dependent dipolar field calculated
for $s$=20~nm. The strength of the initial dipolar field is
$\overline{h_d}$=20.5~Oe, with the direction of the sum vector
along the major axis of the ellipse. The behavior of the
time-dependent dipolar field is similar to the former case up to
800~ps. However, the fluctuation of the field is smaller than in
the other case because the particle separation is larger. For
$s$=20~nm the magnetization switches and the dipolar field does
not increase to the value taken before starting the rotation. A
similar behavior is also found for $s$=30~nm (star line). In that
case $\overline{h_d}$=16.4~Oe and the fluctuation is again
reduced. In Fig~\ref{time-dependent_dipolar_field}b we plot the
variation of the direction of the dipolar field as a function of
time. The direction fluctuations decrease with separation
"\hspace{-1 mm}\textit{s}", like for the strength. In the
following we investigate in more details the case with particle
separation $s$=20~nm. The dot-line in
Fig~\ref{time-dependent_dipolar_field}a depicts the relaxation
behavior of the central particle magnetization in the presence of
the dipolar field without external magnetic field. The dipolar
field increases according to the relaxation of the system from the
starting state into the energetic minimum. After that, the dipolar
field does not change anymore because no external field is
applied. In contrast, by applying the switching field the
magnetization rotates acting on the neighboring particles and,
thus, modifying the time-dependent dipolar field (see full-line).
Finally, the contribution of the motion of the half-selected
particles is evident by comparing the full-dot line with the full
line. In particular, taking this motion into account this leads to
a reduction of the strength of the time dependent dipolar field.
This reduction has its maximum of 7~Oe at t=800~ps.

\begin{figure}\center{\includegraphics[width=10cm]{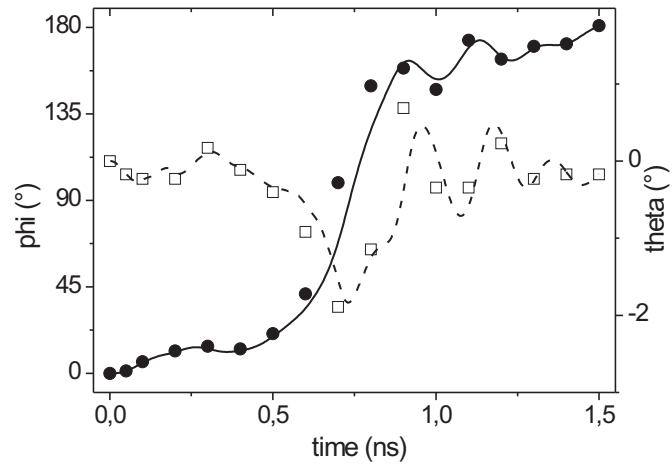}}
\caption{Magnetization trajectory in the presence of the dipolar
field. The trajectory refers to the fully-selected particle in
Fig.~\ref{dipolar_configuration} and it is calculated in the
presence of either the exact dipolar field (full dot, open square
for the components) or the average dipolar field (full and dot
line).} \label{dipolar_trajectory}\end{figure}

In order to simplify the calculation of the switching process and
to shorter the simulation time, it is necessary to approximate the
spatial and time dependent dipolar field with the averaged dipolar
field. Such an approximation drops the simulation time by about a
factor of 10. To justify the use of this approximation the
difference between the switching processes obtained with the
averaged dipolar field and the exact dipolar field has to be
small. In Fig.~\ref{dipolar_trajectory} we plot the
time-dependence of the components of the average magnetization for
the switching process for $s$=20~nm. The trajectory obtained in
presence of the exact dipolar field (full-dot line in
Fig.~~\ref{time-dependent_dipolar_field}a) is represented by the
full dots and the open squares. In contradistinction, we show with
the full and dot lines the components of the average magnetization
obtained with the averaged dipolar field. The trajectories behave
similarly. The main difference is found during the switching
period at t=700~ps signified by a shift of 50~ps between the two
trajectories. However, the final states and the majority of points
during the switching process coincide. Therefore the use of the
averaged dipolar field approximation is justified. Note that for
$s$=20~nm the criterion of multi-switching stability is not
satisfied because the particle of interest in some of the 32
possible magnetic configurations does not switch. In order to
satisfy the criterion the separation has to be larger. To
conclude, we have analyzed the difference between the two
approaches as a function of the separation and verified that the
larger the particle separation the more justified is the use of
the averaged dipolar field approximation.

\begin{center}
\large\textbf{B. Effect of the dipolar interaction on the
switching stability}\normalsize
\end{center}

\begin{figure}\center{\includegraphics[width=10cm]{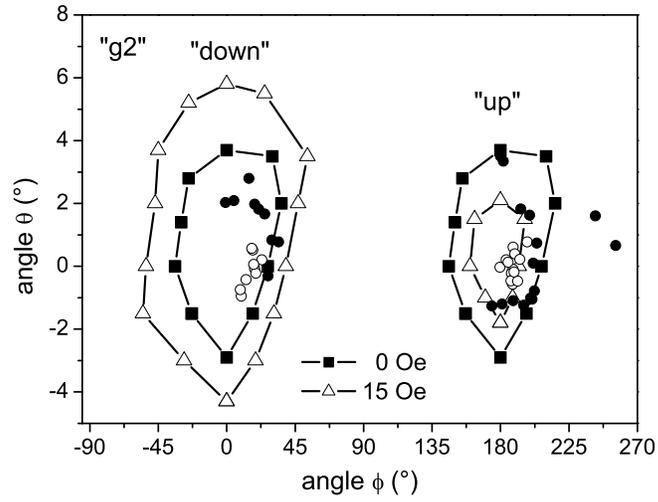}}
\caption{Regions of stability for the addressing scheme "g2".
Square line: regions calculated in absence of the dipolar
interaction. Triangle line: regions calculated in the presence of
the averaged dipolar field $\overline{h_d}$=15~Oe. Full dots:
final states obtained after application of a pulse with duration
$t_p$=0.7~ns from the "triangle" points. Open dots:  final states
obtained after application of a pulse with duration $t_p$=1.5~ns
from the "triangle" points.} \label{dipolar_g2}
\end{figure}

\begin{figure}\center{\includegraphics[width=10cm]{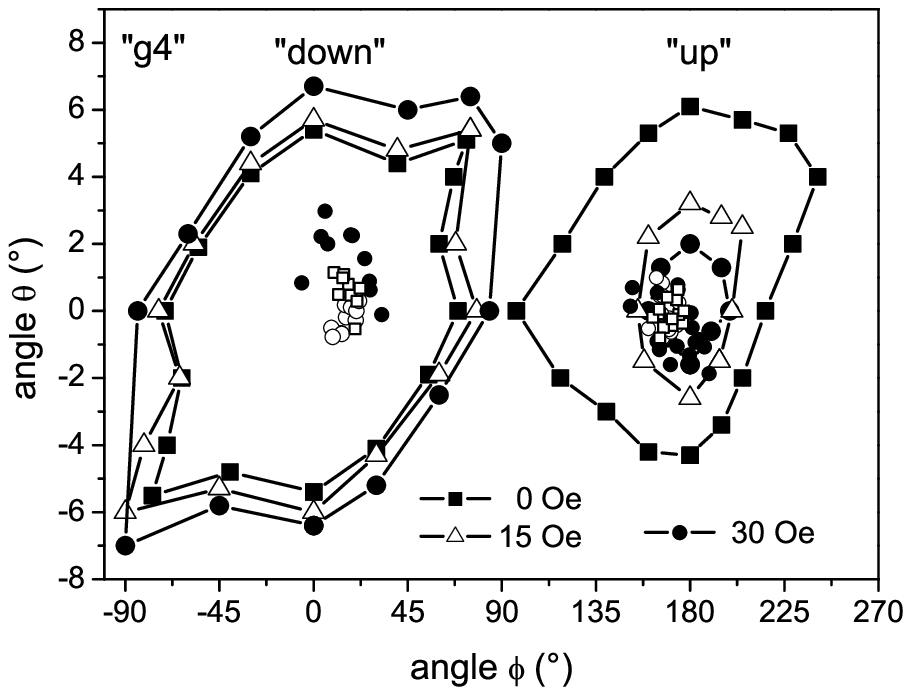}}
\caption{Regions of stability for the addressing scheme "g4".
Square line: regions calculated in absence of the dipolar
interaction. Triangle line: regions calculated in the presence of
the averaged dipolar field $\overline{h_d}$=15~Oe. Full-dot line:
regions calculated in the presence of the average dipolar field
$\overline{h_d}$=30~Oe. Full dots: final states obtained after
application of a pulse with duration $t_p$=0.7~ns from the
"triangle" points. Open dots: final states obtained after
application of a pulse with duration $t_p$=1.5~ns from the
"triangle" points. Open squares: final states obtained after
application of a pulse with duration $t_p$=1.5~ns from the "dot"
points.} \label{dipolar_g4}
\end{figure}

In the following section we study the effect of the dipolar
interaction on the regions of stability and on the magnetization
trajectories for particles subjected to the switching schemes
"g2", "g3" and "g4". According to the results of the previous
section we consider particles subjected to an averaged dipolar
field calculated by taking into account the first and second
neighbors. In order to calculate the averaged dipolar field we
assume the particles to be placed on a square grid, which is the
usual geometry present in the literature. In absence of the
dipolar interaction the particles with thickness $t=2$~nm satisfy
the criterion of multi-switching stability both at zero and room
temperature (see section III). To study the effect of the dipolar
interaction at room temperature we consider the same particle
arrangement subjected to magnetic field pulses with strength
$H_{sw}=130$~Oe. As mentioned in sec. IIA, the classical switching
scheme "g1" does not guarantee sufficient stability in the
presence of the dipolar interaction\cite{maffitt}. According to
our procedure, in order to judge the stability for the switching
scheme "g1", we have to prove the stability of the "up" and "down"
states under the action of the field
\textbf{\textit{h${_\textbf{b}}$}} ($h{_b}$=122.2~Oe), which is
the relevant component of the field
\textbf{\textit{h${_\textbf{sw}}$}}. In absence of the dipolar
interaction we find that an offset $\Delta\theta=0.5^{\circ}$ from
the minimum ($\theta=0^\circ$,$\phi=180^\circ$) is sufficient to
cause instability. This value is much smaller than those obtained
for the switching schemes "g2", "g3" and "g4" even in the presence
of dipolar interaction, see
Figs.~\ref{dipolar_g2}-\ref{dipolar_g4}. Furthermore, the
magnetization placed in the minimum
($\theta=0^\circ$,$\phi=180^\circ$) becomes unstable by adding to
the field \textbf{\textit{h${_\textbf{b}}$}} a magnetic field
$h_{x}$=4~Oe used to simulate the averaged dipolar field. These
examples show that the switching scheme "g1" is inadequate with
regard to stability in the presence of the dipolar interaction. We
now consider switching scheme "g2"\cite{hillebrands}. In order to
find the regions of stability the magnetization is subjected to
the field \textbf{\textit{h${_\textbf{a}}$}}, see
Fig.~\ref{schemes}b, which creates a larger torque than the field
\textbf{\textit{h${_\textbf{b}}$}} and, thus, is the relevant
component of \textbf{\textit{h${_\textbf{sw}}$}} to study the
stability. In Fig.~\ref{dipolar_g2} we plot the regions of
stability calculated for a step function with rise-time
$t_{r}$=75~ps and strength $h_{sw}=130$~Oe. Since the magnetic
torque $|$\textbf{M$\times$\textbf{\textit{h${_\textbf{sw}}$}}}$|$
is invariant for $\textbf{M}$ in the "up" or in the "down" state,
see sec. II-A, in absence of dipolar interaction the regions of
stability are the same. This symmetry is broken in the presence of
the dipolar interaction. For the distance $s=130$~nm, the average
dipolar field produced on the central particle by the first and
second neighbors is $\overline{h_{d}}$=15~Oe, see
Fig.~\ref{dipolar_decay}. Note that this number refers to the
"key"-configuration to study the dipolar coupling, see above in
sec. IV-A. Furthermore, for smaller "\hspace{-1 mm}\textit{s}" the
averaged dipolar field would inhibit a functioning of the
switching scheme "g2". Since \textbf{\textit{h${_\textbf{d}}$}} is
oriented in the "down" direction, the stability of the "down"
state increases, while the one of the "up" state decreases. As a
consequence, the area of the stability regions grows or shrinks
correspondingly, see Fig.~\ref{dipolar_g2}. In order to study the
switching properties, we compare two different pulses with
duration 0.7~ns and 1.5~ns. The pulses are characterized by (0.7,
0.075, 0.1, 0.2, 0.075) and (1.5, 0.075, 0.2, 0.8, 0.075). In
absence of the dipolar interaction, for $t_p=0.7$~ns some of the
trajectories do not terminate inside the stability regions after
switching. On the other hand, for $t_p=1.5$~ns all the
trajectories terminate inside the stability regions and the
criterion of multi-switching stability is satisfied. In
Fig.~\ref{dipolar_g2} we plot the final states obtained after
switching in the presence of the dipolar interaction for
$\overline{h_{d}}$=15~Oe. In particular, we show with the full
dots the states associated to the pulse with duration $t_p=0.7$~ns
and with the open dots the states associated to the pulse with
duration $t_p=1.5$~ns. First we consider the switching from the
"up" state into the "down" state. From Fig.~\ref{dipolar_g2} we
see that all the trajectories terminate inside the stability
region (triangle line). Furthermore, we notice that the dispersion
of the final states increases by decreasing the duration of the
pulse. By considering the switching from the "down" into the "up"
state we observe that some of the final states do not end up in
the stability region. This happens for both pulse durations.
Therefore for the switching scheme "g2" and for the pulses
considered, the criterion of multi-switching stability is not
satisfied for $\overline{h_{d}}$=15~Oe.

We now consider switching scheme "g4" and, according to the
procedure presented in section II-B, we calculate the regions of
stability for a step function with rise-time $t_r$=75~ps and
strength $h_{sw}$=130~Oe. In Fig.~\ref{dipolar_g4} we plot the
resulting regions calculated in absence of the dipolar interaction
and in the presence of the averaged dipolar fields
$\overline{h_d}$=15~Oe and $\overline{h_d}$=30~Oe. According to
Fig.~\ref{dipolar_decay} these fields correspond to arrays of
particles with separations $s=130$~nm and $s=82$~nm, respectively.
Similar to the result obtained for the switching scheme "g2", the
effect of the dipolar coupling is to increase the stability of the
"down" state and to decrease the stability of the "up" state.
Since the "up"-state area is larger than the "down"-state area,
the "up" state mainly determines the stability of the system. In
that sense it is interesting to compare the stability regions of
the "up" state for the switching schemes "g2" and "g4". By
comparing Fig.~\ref{dipolar_g2} and Fig.~\ref{dipolar_g4} we
notice that the area of the stability region of the "up" state
obtained with scheme "g2" for $\overline{h_d}$=15~Oe is similar to
the one obtained with scheme "g4" for $\overline{h_d}$=30~Oe.
Roughly this means that the same results as obtained for the
switching scheme "g2" for an array of particles with separation
$s=130$~nm are obtained for the scheme "g4" for an array of
particles with $s=82$~nm. In order to study the switching
properties, we compare two pulses with duration 0.7~ns and 1.5~ns.
In both cases, in absence of dipolar coupling, all the
trajectories terminate inside the stability regions and the
criterion of multi-switching stability is satisfied. In the
presence of the dipolar interaction all the trajectories from the
"up" state terminate in the "down" state (see
Fig.~\ref{dipolar_g4}). For $\overline{h_d}$=15~Oe with
$t_p=1.5$~ns all the trajectories from the "down" state terminate
in the "up" state. Again, the criterion of multi-switching
stability is satisfied. On the other hand, for
$\overline{h_d}$=15~Oe with $t_p=0.7$~ns and for
$\overline{h_d}$=30~Oe with $t_p=1.5$~ns, some of the trajectories
do not terminate inside the corresponding stability regions.
Therefore for the switching scheme "g4" and for the pulse
considered, the criterion of multi-switching stability is not
satisfied for $\overline{h_{d}}$=30~Oe.

\begin{figure}\center{\includegraphics[width=10cm]{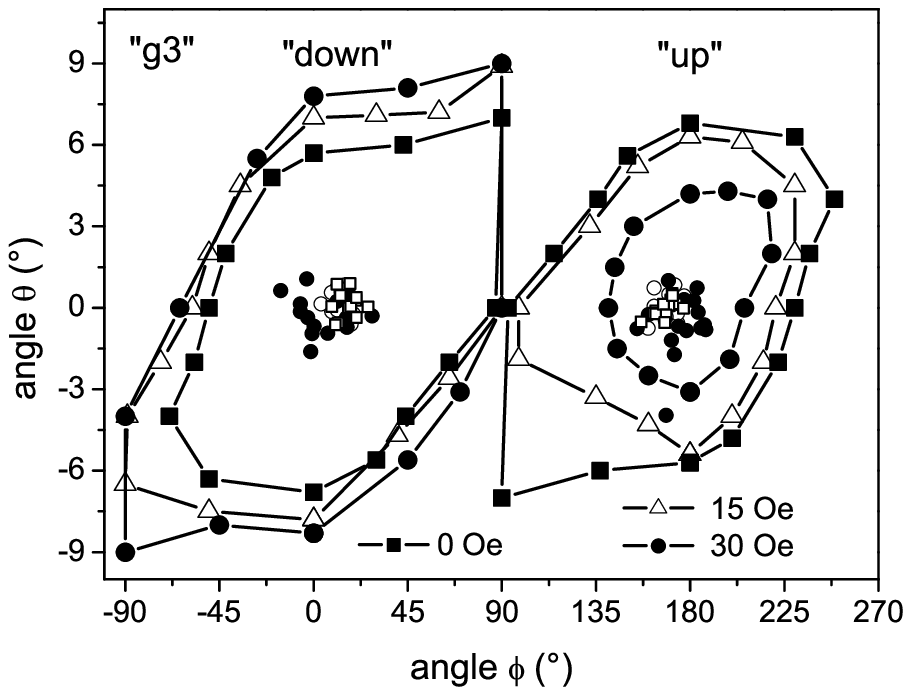}}
\caption{Regions of stability for the addressing scheme "g3".
Square line: regions calculated in absence of the dipolar
interaction. Triangle line: regions calculated in the presence of
the average dipolar field $\overline{h_d}$=15~Oe. Full-dot line:
regions calculated in the presence of the average dipolar field
$\overline{h_d}$=30~Oe. Full dots: final states obtained after
application of a pulse with duration $t_p$=0.7~ns from the
"triangle" points. Open dots: final states obtained after
application of a pulse with duration $t_p$=1.5~ns from the
"triangle" points. Open squares: final states obtained after
application of a pulse with duration $t_p$=1.5~ns from the "dot"
points.} \label{dipolar_g3}
\end{figure}

Finally we consider switching scheme "g3". The only difference
with scheme "g1" is the double line "$b_1$" and "$b_2$" used to
improve the stability of the half-selected particles. The magnetic
field used to calculate the stability regions is
$h_{b1}$=$h_{b2}$=61.1~Oe. In Fig.~\ref{dipolar_g3} we plot the
stability regions obtained in absence of dipolar coupling and in
the presence of the averaged dipolar fields $\overline{h_d}$=15~Oe
and $\overline{h_d}$=30~Oe. By comparing Fig.~\ref{dipolar_g2} and
Fig.~\ref{dipolar_g3} we notice that the area of the stability
region of the "up" state obtained for the scheme "g2" in absence
of dipolar coupling is similar to the one obtained for the scheme
"g3" for $\overline{h_d}$=30~Oe. Therefore by using the switching
scheme "g3" we expect to have an improvement of stability even
higher than with the one obtained by using the switching scheme
"g4". Such an expectation is confirmed by the study of the
switching trajectories. Indeed we find that in the presence of
dipolar interaction for $\overline{h_d}$=15~Oe with $t_p=0.7$~ns
and $t_p=1.5$~ns and for $\overline{h_d}$=30~Oe with $t_p=1.5$~ns
all the trajectories terminate inside the stability regions, for
switchings from the "up" into the "down" states and vice-versa.
Note also that for all these cases the criterion of
multi-switching stability is satisfied.

\begin{center}
\large\textbf{V. Discussion and conclusion}\normalsize
\end{center}

In this paper we have presented a procedure to study the stability
and the switching behavior for an array of magnetic particles. In
the literature studies of this type are usually performed with the
support of the magnetic astroid. In that case the emphasis is put
on the switching aspect of the system. After switching the
magnetization is located in the energetic minimum. In the damped
regime small misalignments from the minimum do not play an
important role for the stability of the system. Therefore in this
regime the exact location of the magnetization after switching is
not taken into account. Rather, other factors, like the not
perfectly equal shape of the particles, the temperature and the
dipolar interaction, are decisive to determine the operation
margin of the astroid, i. e., the range of stability of the
system. On the other hand, in fast dynamics the misalignments
dramatically affect the stability of the system (sec. II-B).
Therefore it is natural to emphasize the study of the stability of
the magnetization, before investigating the switching behavior of
the system (II-B). That is what we have done in this paper. First,
we have studied the stability of the half-selected particles.
Second, we have investigated the switching behavior of the
fully-selected particle. This procedure led us to define the
\textit{criterion of multi-switching stability} which has to be
satisfied in order to have stable and repeatable switching (II-C).
One advantage of this procedure is to give a graphical
representation of the intrinsic stability of the system. Since the
stability of the system directly depends on the area of the
stability regions, to improve the stability one has to enlarge
these regions. It is interesting to note that in the literature
there are two methods to obtain stable switching in the dynamical
regime. One is to prolong the switching time to the time-scale of
the damped switching, allowing the magnetization to relax into the
energy minimum (magnetization ringing). The other one is to shape
the applied magnetic field pulse so that the magnetization reaches
directly the minimum without damped relaxation (ballistic
precessional switching)\cite{miltat,gerrits-schumacher}. Both
methods do not improve the intrinsic stability of the system, i.
e., the stability regions do not grow. Comparing now different
switching schemes, in this work, in accordance with the
literature, we find that the classical addressing scheme "g1" is
not suitable to obtain stable switching. On the other hand, we
have shown three alternative schemes, one of which is known from
the literature ("g2"), which are suitable to obtain stable
switching. The common goal of these schemes is to enlarge the area
of the stability regions, either by changing the orientation of
the conduction line ("g2") or by adding a third conduction line
("g3" and "g4"). Each of these schemes is preferable for some
reason. The switching scheme "g2" works with only two sets of
conduction lines. The scheme "g3" is the most stable in the
presence of dipolar interaction. The scheme "g4" is the only one
which uses only unipolar magnetic field pulses. Nevertheless, each
of these schemes satisfies the criterion of multi-switching
stability at room temperature in the presence of dipolar
interaction and for pulses shaped according to CMOS specifications
(III, IV-B).

To conclude, we have proved that in an array of magnetic particles
it is possible to obtain stable and repeatable switching in the
GHz regime. This result has been obtained by improving the
stability of the half-selected particles' magnetization. In
accordance with other studies of magneto-dynamics in the fast
regime, the magnetization switches by precession. The pulse
shaping used in ballistic precessional switching with the aim to
exactly "shoot" the minimum is not required. The magnetization
trajectories terminate inside the region of stability, without the
necessity to reach a precise point. The switching is precessional,
but non-ballistic. We call this \textit{non-ballistic precessional
switching}.

\begin{center}
\large\textbf{VI. Acknowledgment}\normalsize
\end{center}

The authors thank M. R. Scheinfein for extending his LLG
Micromagnetic Simulator\cite{scheinfein} to include the position
dependent external field tool needed for the present work.

\newpage

\newpage

\begin{thebibliography}{99}
{\footnotesize\markboth{Bibliograbhy}{Bibliography}

\bibitem{miltat} J. Miltat, G. Albuquerque, and A. Thiaville, Top.
Appl. Phys. {\bf 83}, 1 (2002).

\bibitem{bertotti} G. Bertotti, I. Mayergoyz, C. Serpico, and M.
Dimian, J. Appl. Phys. {\bf 93}, 6811 (2003).

\bibitem{bauer} M. Bauer, J. Fassbender, and. B. Hillebrands,
Phys. Rev. B {\bf 61}, 3410 (2000).

\bibitem{kaka} M. Bauer, R. Lopusnik, J. Fassbender, and B.
Hillebrands, Appl. Phys. Lett. {\bf 76}, 2758 (2000); S. Kaka and
S. E. Russek, Appl. Phys. Lett. {\bf 80}, 2958 (2002).

\bibitem{maffitt} T. M. Maffitt, J. K. DeBrosse, J. A. Gabric, E.
T. Gow, M. C. Lamorey, J. S. Parenteau, D. R. Willmott, M. A. Wood
and W. J. Gallagher, IBM J. Res. Dev. {\bf 50}, 25 (2006).

\bibitem{engel} B.N Engel, J. Akerman, B. Butcher, R. W.  Dave,
M. DeHerrera, M. Durlam, G. Grynkewich, J. Janesky, S. V.
Pietambaram, N. D. Rizzo, J. M. Slaughter, K. Smith, J. J. Sun and
S. Tehrani, IEEE Trans. Magn. {\bf41}, 132 (2005).

\bibitem{hillebrands} Patent No.: US 6,674,662 B1; M. Bauer, J. Fassbender,
B. Hillebrands and R. L. Stamps, Phys. Rev. B {\bf 61}, 3410
(2000).

\bibitem{brockmann} M. Brockmann, S. Miethaner, R. Onderka, M.
K\"{o}hler, F. Himmelhuber, H. Regensburger, F. Bensch, T.
Schweinb\"{o}ck, and G. Bayreuther, J. Appl. Phys. {\bf 81}, 5047
(1997).

\bibitem{scheinfein} See http://llgmicro.home.mindspring.com

\bibitem{thiaville} A. Thiaville, Phys. Rev. B {\bf61}, 12221
(2000).

\bibitem{maunoury} C. Maunoury, T. Devolder, C. K. Lim, P. Crozat,
and C. Chappert, J. Appl. Phys. {\bf 97}, 074503 (2005).

\bibitem{schumacher2} H. W. Schumacher, Appl. Phys. Lett. {\bf 87}, 042504
(2005).

\bibitem{garcia} J. L. Garcia-Palacios and F. J. Lazaro, Phys.
Rev. B {\bf22}, 14937 (1998).

\bibitem{deak} J. G. Deak, J. Appl. Phys. {\bf 93}, 6814 (2003).

\bibitem{cowburn} R. P. Cowburn, J. Phys D {\bf 33}, R1 (2000).

\bibitem{bromwich} T. J. Bromwich, A. Kohn, A. K. Petford-Long, T.
Kasama, R. E. Dunin-Borkowski, S. B. Newcomb and C. A. Ross, J.
Appl. Phys. {\bf 98}, 053909 (2005).

\bibitem{fab1} F. Porrati and M. Huth, J. Magn. Magn. Mater. {\bf 290}, 145 (2005).

\bibitem{guslienko} K. Yu Guslienko, Appl. Phys. Lett. {\bf 76},
3609 (2000).

\bibitem{fab2} F. Porrati and M. Huth, Appl. Phys. Lett. {\bf 85},
3157 (2004).

\bibitem{dunin} R. E. Dunin-Borkowski, M. R. McCartney, B. Kardynal, D. J. Smith,
M. R. Scheinfein, Appl. Phys. Lett. {\bf 75}, 2641 (1999).

\bibitem{abraham} M. C. Abraham, H. Schmidt, T. A. Savas, H. I. Smith,
C. A. Ross, R. J. Ram, J. Appl. Phys. {\bf 89}, 5667 (2001).

\bibitem{devolder} T. Devolder and C. Chappert, J. Appl. Phys. {\bf 95}, 1933
(2004).

\bibitem{pham} H. N. Pham, I. Dumitru, A. Stancu and L. Spinu,
J. Appl. Phys. {\bf 97}, 10P106 (2005).

\bibitem{gerrits-schumacher} T. Gerrits, H. A. M. van den Berg,
J. Hohlfeld, L. B\"{a}r, and T. Rasing, Nature {\bf 418}, 509
(2002); H. W. Schumacher, C. Chappert, P. Crozat, R. C. Sousa, P.
P. Freitas, and M. Bauer, Appl. Phys. Lett. {\bf 80}, 3781 (2002);
H. W. Schumacher, C. Chappert, R. C. Sousa, P. P. Freitas, and J.
Miltat, Phys. Rev. Lett. {\bf 90}, 017204 (2003)


}
\end{thebibliography}
\end{document}